\documentstyle[preprint,aps]{revtex}

\tightenlines

\begin{document}

\draft

\title{Convergences in the Measurement Problem in Quantum Mechanics}

\author{L. F. Santos $^1$ and C. O. Escobar $^2$}

\address{$^1$ Departamento de F\'{\i}sica Nuclear \\
Instituto de F\'{\i}sica da Universidade de S\~ao Paulo, 
C.P. 66318, cep 05389-970 \\
S\~ao Paulo, S\~ao Paulo, Brazil\\
lsantos@charme.if.usp.br \\
$^2$ Departamento de  Raios C\'osmicos e Cronologia \\
Instituto de F\'{\i}sica Gleb Wataghin \\
Universidade Estadual de Campinas, C.P. 6165,  cep 13083-970\\
Campinas, S\~ao Paulo, Brazil\\
escobar@ifi.unicamp.br}

\maketitle

\begin{abstract}

This paper presents arguments purporting to show
that von Neumann's
description of the measurement 
process in quantum mechanics has a modern day version in
the decoherence approach. We claim that this approach and
the de Broglie-Bohm theory
emerge from Bohr's interpretation and are therefore 
obliged to deal with some obscure ideas
which were anticipated, 
explicitly or implicitly and carefully circumvented, by Bohr.

\end{abstract}

\section{Introduction}

The existence of quantum physics imposes on physicists an unavoidable 
ambiguity when describing atomic and subatomic systems. On his/her daily 
activity in the laboratory, the physicist describes and calculates 
classical trajectories of particles in order not only to interpret the data, 
but also to design the apparatus producing that data. Witness, 
for example, how a 
high energy physicist reconstructs events involving a maze of particles 
produced in a complex subatomic collision. He/She rarely uses any 
quantum mechanics at all, determining trajectories, lifetimes, vertex 
positions, trigger algorithms, all with classical relativistic mechanics.
Bohr was perfectly aware of this necessity of using classical language 
for describing the results of experiments and constructed his interpretation 
of quantum mechanics based on this 
(Bohr 1939). The need for a 
classical language is imposed, according to him, 
by the classical nature of observers and experimental 
apparatuses. It is fair to say that for Bohr there 
was no measurement problem, as 
classical apparatuses are not described by 
wave functions, avoiding superpositions of 
macroscopic states. Wave functions pertain only to the microscopic 
world. The problem as it is recognized today can 
be traced back to von Neumann.

We argue in this paper that von Neumann's interpretation of quantum 
mechanics (von Neumann 1955) originated in an attempt to remove the somewhat
arbitrary division between the classical and the quantum world 
introduced by Bohr. In so
doing, von Neumann shifted the cut by introducing an observer, who
was not required by Bohr.

Modern attempts to solve the measurement problem introducing the 
environment to dissipate macroscopic coherence do not explain the 
collapse of the wave function. We argue in the text that 
decoherence models (Zurek 1998) 
are true descendants of von Neumann and therefore 
will ultimately bring the observer to the forefront. Consequently, the 
decoherence approaches are not a solution of the measurement problem if 
one's standing point is that the observer should not play a role in the 
interpretation.

We further argue that the causal approach, introduced by de Broglie and 
developed by Bohm and collaborators 
(Bohm 1952 and 1995), also originated in an effort to better 
deal with the division of the world. The causal approach removed it 
by combining classical and quantum concepts 
in a single description of nature: from the classical world it takes 
the position of the particles (be they part of the system or 
of the measuring device), while keeping from the quantum world the 
wave function and its Schr\"odinger's evolution.

We also bring forward our view that von Neumann's (and its 
modern day version - decoherence) and the de Broglie-Bohm interpretation,
though corresponding to two different branches emerging from 
Bohr's elaborate world view, dealt in their own 
specific way with the vague concept of information and the quite obscure
notion of its disappearance.

\section{Bohr}

With a long historical hindsight, we can now see Bohr's position as one 
that intended to provide an interpretation, whose main 
purpose was to protect the successful formalism of quantum mechanics. 
We might say that Bohr anticipated many of the problems that would be faced 
by those who would later try to analyze in detail the measurement 
process. As we will show below, the attempts presented here 
to solve the measurement have to answer questions that 
do not pertain to the daily activity 
of an experimenter in the laboratory. Bohr somehow foresaw these 
inextricable difficulties and cut them short by declaring (Bohr 1939):

`In the system to which the quantum mechanical formalism is applied, 
it is of course possible to include any intermediate auxiliary 
agency employed in the measuring process [but] some ultimate 
measuring instruments must always be described entirely 
on classical lines, and consequently kept outside the system subject 
to quantum mechanical treatment.'

Although Bohr's position was a strong and deeply intricate one, it  
was challenged by one simple criticism: where is the demarcation between 
system and apparatus, quantum and classical?

`The `Problem' then is this: how exactly is the world 
to be divided into speakable apparatus ... that we can talk about ... 
and unspeakable quantum system that we can not talk about? How many 
electrons, or atoms, or molecules, make an `apparatus'? The 
mathematics of the ordinary theory requires such a division, but 
says nothing about how it is to be made. 
In practice the question is resolved by pragmatic 
recipes which have stood the test of time, applied with discretion 
and good taste born of experience. But should not fundamental 
theory permit exact mathematical formulation?' (Bell 1987, 171)

Though simple, this question has a devastating effect on 
Bohr's interpretation. This quotation from Bell summarizes the 
challenge and motivation for those who felt the urge to explain 
the measurement process despite the best 
advice against it by Bohr.

\section {Von Neumann}

Von Neumann was probably the first to attempt a unified quantum description 
of system and apparatus. Contrary to Bohr, who avoided the danger 
of such a description by constructing a philosophical fortress 
around quantum mechanics, von Neumann made a formal analysis of the 
measurement process and ended up by arriving at an altogether 
new interpretation of quantum mechanics, which no wonder is 
frequently misidentified with Bohr's philosophical constructs. 
As stressed by Feyerabend (Feyerabend 1962, 237):

`when dealing with von Neumann's investigation, we are not dealing 
with a refinement of Bohr - we are dealing with a completely different 
approach.' 

Like Bohr, von Neumann attributed importance to the apparatus as 
part of the measurement process, but he examined the evolution 
of the joint system (system + apparatus) with a single wave function 
governed by Schr\"odinger's evolution, which establishes a correlation 
between them in such a way that any result pertaining to the system 
is inferred from the reading of the apparatus.

In doing so, the states of the apparatus are also subjected to the 
superposition principle. Clearly von Neumann managed to move the 
classical/quantum cut from the system/apparatus boundary, 
but at the price of leaving the apparatus in a coherent superposition 
of states which is not observed. No matter how many 
apparatuses are included, the superpositions will remain. At this 
stage von Neumann distinguished two types of processes in quantum mechanics: 
the one described above, leading to undesirable macroscopic 
superpositions as a consequence of the {\it reversible} 
unitary evolution of 
Schr\"odinger's equation and the other one, 
corresponding to our knowledge of the result of the measurement, which is 
{\it irreversible}. Following Bohr,

`it is also essential to remember that all unambiguous 
information concerning atomic object is derived from the permanent marks... 
left on the bodies which define the experimental conditions. 
Far from involving any special intricacy, the irreversible amplification 
effects on which the recording of the presence of atomic objects rests 
rather remind us of the essential irreversibility inherent in 
the very concept of observation.' (Bohr 1964, 3)

Von Neumann formalized the irreversibility in quantum mechanics 
by postulating the collapse of the wave function. Notice, though,
that he deals with ensembles and therefore uses 
density matrices in his formalism. To avoid imposing the postulate 
without any physical justification, the observer is introduced and 
his/her subjective perception becomes essential. 
This interpretation is thereby weakened and open to 
severe criticisms. The cut is still present, but has now moved to a 
position between joint system/observer.

In addition to interpretative and epistemological problems, this 
interpretation also has problems in its formalism. 
The need to consider instantaneous interactions in the 
measurement process, so that the unitary evolution does not move the 
state vector away from the position of measurement, implies that 
the Hamiltonian for the joint system commutes with the observable 
which is being measured $[H, O] = 0$. This could be a demanding 
condition on the Hamiltonian, but not an excessive one, as we will 
make clear when discussing decoherence.

\section{Decoherence}

Von Neumann's approach is taken one step further in the decoherence 
models. These models invoke the inevitable interaction between 
joint system and environment to help solving, it is claimed, the 
measurement problem. Following von Neumann's tradition, system, apparatus 
and environment are treated quantum mechanically and, as for  
von Neumann, the unavoidable superposition of macroscopically different 
states will still be present. As the environment has a large number of 
degrees of freedom, the observer has no access to them and therefore, 
they must be traced over, ignored. 
Notice that the observer still plays a crucial 
role in this approach, for the trace must be done by someone 
and the cut is maintained as the boundary between the degrees of 
freedom which are traced over and those which are not. The inevitability
of this division of the world is
acknowledged by the proponents of this approach as 
illustrated by Zurek in a recent
paper (Zurek 1998, 1794):

`We can mention two such open issues right away: both the 
formulation of the measurement problem and its resolution through 
the appeal of decoherence require a universe split into 
systems.'

The trick of tracing over the unaccessible degrees of freedom 
brings the density matrix  of the total system to a diagonal form 
removing the undesirable macroscopic superpositions and this is 
von Neumann's postulate presented in a more  elaborate dynamical way.
However, there is a subjective element in the whole procedure: 
how far does the environment reach?

Von Neumann's condition of commutativity of the Hamiltonian and the 
observable to be measured, now acquires a  
complex meaning: $[H_{int} ,A]$, where now $A$ is the pointer 
basis of the apparatus and 
the Hamiltonian refers to the interaction 
between joint system and environment, for which there is even 
less control by an experimental physicist. It is as if a
measuring  instrument should bring in its instruction manual 
recommendations on the appropriate environment where to operate. 
Von Neumann's condition only indicated the adequate 
apparatuses for measuring a certain observable. 
Demands put on the experimenter do not stop here however, he/she - 
according to the decoherence procedure known as the predictability 
sieve (Zurek 1993) - should refer to a list to decide which observable 
can be measured, those on the top of the list are more classical in 
appearance and thus preferable for measurement.
The existence of such a list brings back the subjectivity in the choice 
of where to put the quantum/classical cut, which is 
`to be decided by circumstances' (Zurek 1993). Besides, how is 
it possible to know the environment well enough to decide which 
observable can be measured, but at the same time to be so ignorant 
about it that one is obliged to trace it over?

One of the worst aspects of the decoherence approach, taken to its 
ultimate consequence, is thus to introduce a set of procedures that 
should be obeyed when measuring a quantity, 
procedures which no experimentalist 
on his/her right mind would recognize as what goes on in the laboratory.
The appeal to notions far remote from the reality of the laboratory 
experiments is  well illustrated in the following passages in a recent paper 
on decoherence (Zurek 1998, 1796 and 1799).

`Correlations [between states of the joint system and environment] are 
both the cause of decoherence and the criterion used to evaluate the 
stability of the states...Moreover, stability of
the correlations between the states of the system monitored by their 
environment and of some other `recording' system (i.e. an apparatus 
or a memory of an observer) is a criterion of the `reality' of these 
states.' 

or still,

`the observer can know beforehand what (limited) set of observables 
can be measured with impunity. He will be able to select 
measurement observables that are already monitored by the 
environment.' 

The above passages clearly show some subjective elements of this approach,
invoking memory of observer or a priori knowledge of the interaction 
between the environment and joint system.

One last criticism is that decoherence, as well as von 
Neumann, deals with the density matrix, which is by force in the realm 
of the ensemble interpretation and as Bell says:

`If one were not actually on the look-out for probabilities, I 
think the obvious interpretation of even [the butchered 
density matrix] would be that the system is in a state in which the various 
[wave functions] somehow co-exist...This is not at all a 
probability interpretation, in which the different terms are seen not as 
co-existing, but as alternatives.' (Bell 1990, 36 [ref.2] 
and Whitaker 1996, 289) 

\section{de Broglie - Bohm}

A description of individual events was proposed by de Broglie-Bohm 
(Bohm 1952). In it, an individual system is described by a wave function 
and a particle. The particle is guided by the wave function, which 
works, to a certain extent, like a field. One could say that this 
theory corresponds to a refinement of Bohr's duality - the use of 
quantum concepts in a scale and classical concepts in another one - taken 
to an extreme. The wave function and the particle position are now 
used at the same time in all scales. It thus eliminates the 
division quantum/classical and apparently has no measurement problem,
as the particle has always a definite position. 
Moreover, it claims to be free from problems connected with the
act of observation, contrary to what we will suggest below.

It is arguably that this theory is subjected to 
some serious criticisms (Holland 1993), 
but the only one we want to emphasize here is related to the infamous
empty wave and more specifically, to the information carried by it. 
Whenever the wave splits up into parts which do not have 
spatial overlap, such as in the trajectories of the double-slit 
experiment, one part will be with the particle and the other one 
will be empty, though it can still influence the particle motion. 
The empty wave 
carries information on the superpositions of states, but as soon as 
a measurement is realized, the empty wave loses any overlap it 
had before with the branch that carries the particle.

`Perhaps we shouldn't talk about it actually disappearing from 
the universe. Rather the information in the `empty' wave 
packet no longer has any effect, because during the act of 
measurement the irreversible process introduces a stochastic 
or random disturbance which destroys the information of quantum 
potential of the wave packet.' (Hiley 1986, 146)

Suddenly the superpositions are destroyed, this only happens thanks 
to the measurement which identifies which branch corresponds to the 
empty wave. How this happens, where 
the empty wave information is taken to, what are the effects 
of its disappearance on the surroundings are left unspecified. 
This sends us back to the similar problem encountered in the decoherence 
models, where information - as before, information on superpositions 
of states - was dissipated into an environment with no observable 
effects on it, but only on the system.

At this point the convergence of these two apparently different 
interpretations becomes clear. If one accepts the concept 
of information as they do, the act of measurement implies its loss, 
be it dissipated in the environment or in the arbitrary 
sterilization of the empty wave, whose information is now 
declared passive:

`we've tried to introduce a distinction between active 
information and inactive information. That is, when an apparatus has 
undergone this irreversible change, one wave packet becomes inactive.'
(Hiley 1986, 146)

This vague notion of information is central to both 
causal and decoherence interpretations. Its vagueness has been nicely 
expressed by Bell (Bell 1990, 8 [ref.3]):

`I don't have a concept of disembodied information - it must 
be located and represented in the material world, and I don't 
know how to formulate the concept of how much information 
there is in an arbitrary space region - I think the concept of 
information is again a very useful one in practice but not 
in principle...'

This concludes our arguments. It is certainly frustrating that 
the measurement problem in quantum mechanics, after six 
decades of being delineated, remains open. Perhaps this indicates 
a fundamentally new epistemological obstacle to be overcome 
jointly by physicists and philosophers, an obstacle that was not present 
in classical physics.

\section{Conclusion}

We argued in this paper that von Neumann's approach (and its modern 
version: decoherence) and the causal interpretation have many points 
in common, despite being 
so different in formalism and in language. 
Remarkably the common points are the problematic ones 
springing from Bohr's deep analysis of the interface quantum/classical. 
These elaborated attempts to define quantitatively what happens in this 
boundary have exposed open unsurmountable problems which Bohr 
carefully avoided by his radical separation of classical and quantum.

\vskip 1cm
\begin{leftline}
{\bf NOTES}
\vskip 0.2cm
*The authors acknowledge the support of the 
Brazilian Research Council, CNPq and would like to thank 
Osvaldo Pessoa Jr. for many discussions on the subject.
\end{leftline}

\end{document}